\documentclass[pra,aps,twocolumn,showpacs,longbibliography]{revtex4-2}
\usepackage{amsmath,amssymb,graphicx}

\usepackage{float}
\usepackage{comment}
\usepackage{bm}
\usepackage{bbold}

\usepackage{graphicx}
\usepackage{dcolumn}
\usepackage{bm}
\usepackage{hyperref}
\usepackage{xcolor}
\usepackage{braket}
\usepackage{textcomp}
\usepackage{comment}
\usepackage{inputenc}

\DeclareMathOperator{\tr}{Tr}

\begin{document}

\title{Improving quantum metrology protocols with programmable photonic circuits}

\author{A. Mu\~{n}oz de las Heras}
\email{alberto.munoz@iff.csic.es}
\affiliation{%
Institute of Fundamental Physics IFF-CSIC, Calle Serrano 113b, 28006 Madrid, Spain}%

\author{D. Porras}
\affiliation{%
Institute of Fundamental Physics IFF-CSIC, Calle Serrano 113b, 28006 Madrid, Spain}%

\author{A. Gonz\'{a}lez-Tudela}
\affiliation{%
Institute of Fundamental Physics IFF-CSIC, Calle Serrano 113b, 28006 Madrid, Spain}%

\date{\today}

\begin{abstract}

Photonic quantum metrology enables the measurement of physical parameters with precision surpassing classical limits by using quantum states of light. However, generating states providing a large metrological advantage is hard because standard probabilistic methods suffer from low generation rates. Deterministic protocols using non-linear interactions offer a path to overcome this problem, but they are currently limited by the errors introduced during the interaction time. Thus, finding strategies to minimize the interaction  time of these non-linearities is still a relevant question. In this work, we introduce and compare different deterministic strategies based on continuous and programmable Jaynes-Cummings and Kerr-type interactions, aiming to maximize the metrological advantage while minimizing the interaction time. We find that programmable interactions provide a larger metrological advantage than continuous operations at the expense of slightly larger interaction times. We show that while for Jaynes-Cummings non-linearities the interaction time grows with the photon number, for Kerr-type ones it decreases, favoring the scalability to big photon numbers. Finally, we also optimize different measurement strategies for the deterministically generated states based on photon-counting and homodyne detection.

\end{abstract}

\maketitle

\section{Introduction} 
\label{sec:Introduction}

Photonic quantum metrology~\cite{DEMKOWICZDOBRZANSKI2015,Dowling2015,Pirandola2018,Polino2020} uses quantum states of light to provide an advantage in the estimation of an unknown parameter $\varphi$ beyond the one achievable with classical resources. The standard configuration for this purpose is the so-called Mach-Zehnder interferometer (MZI) in which two light probes are sent into a beam splitter, after which a phase difference $\varphi$ is encoded between the two paths, which are afterwards mixed again in another beam splitter before measuring some observable. With classical light, the estimation error is always lower bounded by the standard quantum limit (SQL), i.e., $\Delta\varphi>1/\sqrt{N}$, with $N$ being the mean photon number of the state. On the contrary, using quantum states of light, like twin Fock states (TFS) $\ket{N/2,N/2}$~\cite{Holland1993}, one can obtain estimation errors below the SQL, in this case $\Delta\varphi=\sqrt{2/N(N+2)}$. The ultimate precision bound is $\Delta\varphi=1/N$, known as the Heisenberg limit (HL), which can be obtained if NOON states, $\ket{\mathrm{NOON}}=(\ket{N,0}+\ket{0, N})/\sqrt{2}$, come out after the first beam splitter of the MZI.  However, obtaining such a quantum metrological advantage in photonics is still an outstanding challenge. First, because generating metrologically useful quantum states with large photon numbers is hard and, second, because the optimal measurement scheme to attain such an advantage depends on the probe states.

Most popular methods to generate metrologically useful states with large photon numbers are based on combining single photons through post-selection~\cite{Mitchell2004,Afek2010,Xiang2011,Israel2012, Yao2012,Guerlin2007,Deleglise2008,Sayrin2011,Wang2016b,GonzalezTudela2017,Zhong2018,Deng2024NatPhys}. However, they suffer from decreasing efficiency rates with increasing photon numbers. Deterministic protocols based on non-linear interactions~\cite{Vogel1993,Law1996,Hofheinz2008,Hofheinz2009,Kirchmair2013,Gonzalez-Tudela2015,Zhang2018,Li2020,Uria2020,Walschaers2021,Yang2022sequential,Eickbusch2022,Long2022,Uria2023,chen2024bosonic} offer a path to avoid this problem, but suffer from low fidelities due to the errors accumulated during the interaction time. Inspired by recent experimental~\cite{He2023,Cimini2024,nielsen2024} and theoretical~\cite{KRISNANDA2021,Steinbrecher2019,Nigro2022,Scala2024} research, we recently pointed out in Ref.~\cite{MunozDeLasHeras_2024} that optimizing programmable photonic circuits, i.e., combining quenches of photon tunnelings and non-linear interactions, can reduce the number of operations to generate metrologically relevant states with respect to other deterministic protocols~\cite{Vogel1993,Law1996,Hofheinz2008,Hofheinz2009,Kirchmair2013,Gonzalez-Tudela2015,Zhang2018,Li2020,Uria2020,Walschaers2021,Yang2022sequential,Eickbusch2022,Long2022,Uria2023}. However, there were still several open questions: Does the reduced number of operations imply smaller interaction times? What is the nature of the generated states? Can the programmable circuits also improve the measurement part?

\begin{figure*}[tb]
    \centering
    \includegraphics[width=\linewidth]{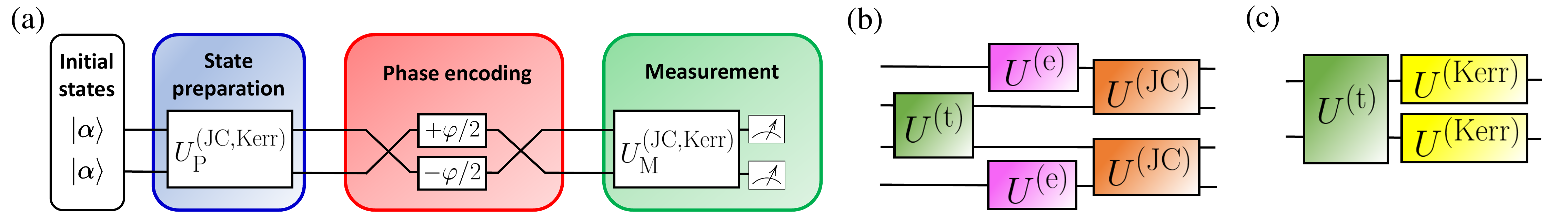}
    \caption{(a) Sketch of the estimation process involving two photonic modes. Two coherent initial states $\ket{\alpha}$ are employed as the input of a state preparation step that involves the unitary operator $U^{\rm (JC,Kerr)}_{\rm P}$. In the continuous approach, this represents the time evolution under the Jaynes-Cummings (JC) or Kerr Hamiltonian. In the programmable approach, it corresponds to a parametrized quantum circuit (PQC) determined by an optimization loop aimed at maximizing the quantum Fisher information (QFI). The prepared state is then sent through a Mach-Zehnder interferometer (MZI) where a phase difference $\varphi$ is encoded between the two modes. While in the continuous approach the measurement takes place immediatedly after phase encoding, in the programmable approach one introduces an additional PQC described by the unitary  $U^{\rm (JC,Kerr)}_{\rm M}$ to prepare an optimal measurement. In this step, one aims at maximizing the classical Fisher information (CFI). (b, c) Sketches of the quantum circuits employed as $U^{\rm (JC,Kerr)}_{\rm P}$ and $U^{\rm (JC,Kerr)}_{\rm M}$ in the programmable approach: (b) represents the JC ansatz (where the two inner wires correpond to the photonic modes and the two outer wires are the emitters), while (c) corresponds to the Kerr ansatz.}
    \label{fig:sketch}
\end{figure*}

In this work, we answer these questions by benchmarking the programmable strategy of Ref.~\cite{MunozDeLasHeras_2024} in both the generation and measurement challenges of the photonic quantum metrology problem. We consider programmable photonic circuits with two classes of non-linearities: a Jaynes-Cummings (JC) interaction, like the one appearing in cavity QED~\cite{Haroche2006exploring}, and a Kerr-type interaction~\cite{butcher_cotter_1990}. However, differently from Ref.~\cite{MunozDeLasHeras_2024}, we analyze in detail the more physically relevant interaction time, and also compare this programmable approach with the conceptually simpler continuous time evolution~\cite{Kirchmair2013,Uria2020, Uria2023} under the non-linearity. This allows us to establish when the programmable approach is advantageous, but also to illustrate the origin of the unexplained scaling behavior found in Ref.~\cite{MunozDeLasHeras_2024}. In particular, we show that even though continuous operations can generate probe states featuring a quantum advantage in parameter estimation, programmability can lead to states approaching the HL while requiring similar interaction times. Finally, we also study the effect of optimizing the programmable photonic circuit before two types of measurements: photon counting~\cite{Cheng2023,Eaton2023} and homodyne photon detection~\cite{Mallet2011,Eichler2011,Bergeal2012}, finding an improvement in both cases, although less pronounced for the latter measurement strategy.

The manuscript is organized as follows: In Sec.~\ref{sec:Scheme}, we define the physical setup of photonic quantum metrology, the programmable and continuous strategies that we employ, and the mathematical tools to quantify their metrological advantage. In Sec.~\ref{sec:continuous}, we study the generation of probe states using the continuous unitary time evolution under fixed photonic non-linearities. In Sec.~\ref{sec:Programmable}, we benchmark these results with the preparation of probe states employing programmable photonic quantum circuits with tunable non-linearities. In Sec.~\ref{sec:Measurements}, we analyze the optimization of the measurement part. Finally, our findings are summarized in Sec.~\ref{sec:Conclusions}.

\section{General concepts of photonic quantum metrology} 
\label{sec:Scheme}

Here, we start by reviewing in Sec.~\ref{sec:Fisher} the definition of the classical and quantum Fischer information, which we employ throughout this manuscript to characterize the metrological power of the quantum states of light we generate. Then, we describe the three parts of photonic quantum metrology setups [see Fig.~\ref{fig:sketch}(a)]: probe state preparation (in Sec.~\ref{sec:Preparation}), parameter encoding (in Sec.~\ref{sec:Phase encoding}), and measurement scheme (in Sec.~\ref{sec:Measurement}), explaining in each of them the continuous and programmable non-linear time evolutions. Further information about the numerical simulations employed to calculate the results of our manuscript can be found in Sec.~SM1  of the Supplementary Material (SM). Codes to reproduce the results of this manuscript are available in~\cite{GitHub}.

\subsection{Definition of classical and quantum Fisher information~\label{sec:Fisher}}

The ultimate precision limit of a certain probe state described by a density matrix $\rho$ is quantified by the \textit{quantum Fisher information} (QFI)~\cite{Braunstein1994,Paris2009}, which is independent of the measurement choice. If the interaction between probe and system is unitary, i.e., $\rho_\varphi=e^{-iH\varphi}\rho e^{iH\varphi}$ where $H$ is a Hermitian operator, the QFI is independent of the unknown parameter $\varphi$ and one can efficiently calculate it using the expression~\cite{Paris2009,Polino2020,Beckey2022}
\begin{equation}
    \mathcal{F}_{\rm Q} = 8 \lim_{\delta\rightarrow 0} \frac{1-F[\rho_\varphi,\rho_{\varphi+\delta}]}{\delta^2}\,
\label{eq:QFI}
\end{equation}
being $F(\rho_1,\rho_2)=\tr[\sqrt{\sqrt{\rho_1}\rho_2\sqrt{\rho_1}}]$ the fidelity between states $\rho_1$ and $\rho_2$.

For a particular measurement scheme, the attainable precision is given by the \textit{classical Fisher informaction} (CFI)~\cite{Fisher1922}
\begin{align}
    \mathcal{F}_{\rm C}(\varphi) = \sum_\lambda P(\lambda|\varphi)\Bigg(\frac{\partial\log[P(\lambda|\varphi)]}{\partial\varphi}\Bigg)^2
    ,
\label{eq:CFI}
\end{align}
where $P(\lambda|\varphi)$ is the probability to obtain a measurement outcome $\lambda$ given a parameter value $\varphi$.

The minimum estimation error attainable by a certain state is given by $(\Delta\varphi)^2=1/\mathcal{F}_{\rm Q}$. Similarly, an optimal measurement satisfies $\mathcal{F}_{\rm Q}=\mathcal{F}_{\rm C}$.
Such a hierarchy is summarized by the \textit{quantum Cramer-Rao bound}~\cite{DEMKOWICZDOBRZANSKI2015,Polino2020}
\begin{align}
    (\Delta\varphi)^2\geq\frac{1}{\nu\mathcal{F}_{\rm C}}\geq\frac{1}{\nu\mathcal{F}_{\rm Q}},
\end{align}
where $\nu$ is the number of independent repetitions of the estimation process.

In photonic quantum metrology, an archetypical problem consists on determining an unknown phase difference $\varphi$ between two photonic modes forming the arms of an MZI, see Fig.~\ref{fig:sketch}(a). Let us now explain in detail the different steps of the protocol.

\subsection{Probe preparation} 
\label{sec:Preparation}

For the preparation stage, we consider the application of unitary time evolutions given by two types of photonic non-linearities to easy-to-prepare states. The first type of non-linearity considered is provided by the interaction of each photonic mode with a single two-level emitter, like in the Jaynes-Cummings (JC) model of cavity QED setups~\cite{Shore1993,Grynberg_Aspect_Fabre_2010}, while the other one is a photon-photon Kerr-type non-linearity~\cite{butcher_cotter_1990}. They are described by the following Hamiltonians, respectively:
\begin{align}
\label{eq:interaction JC}
    H^{(\rm JC)} &= g\sum^{2}_{i=1} \left(\sigma^\dagger_i a_i + \sigma_i a^\dagger_i\right)\,,\\
    H^{(\rm Kerr)} &= K\sum^{2}_{i=1}(a^{\dagger}_i a_i)^2\,,
\label{eq:interaction Kerr}
\end{align}
where $g$ and $K$ are the interaction strength in each case, $\sigma_i = \ket{g}_i \bra{e}_i$ is a lowering operator for the emitter interacting with the $i$-th cavity, and $a^{(\dagger)}_i$ are the annihilation (creation) operators for the photonic mode $i$. 

For both non-linearity types, the two photonic modes are initialized in easy-to-prepare states, i.e., two coherent states $\ket{\alpha}\otimes\ket{\alpha}=\ket{\alpha,\alpha}$ with the same mean photon number $|\alpha|^2=N/2$, such that the total mean photon number is $N$. Besides, in the JC case, we consider the emitters initially in their ground states. Thus, the initial states for the two situations read $\ket{\psi^{\rm (JC)}_0}=\ket{g}\otimes\ket{g}\otimes\ket{\alpha,\alpha}=\ket{g,g,\alpha,\alpha}$ and $\ket{\psi^{\rm (Kerr)}_0}=\ket{\alpha,\alpha}$, respectively. 

We study two different ways of applying these non-linearities to the initial states:
\begin{itemize}
    \item Letting the initial coherent states continuously evolve through the unitary dynamics of the non-linear Hamiltonians of Eqs.~(\ref{eq:interaction JC}-\ref{eq:interaction Kerr}). The probe state is then given by $\ket{\psi^{\rm (JC,Kerr)}_{\rm P}} = e^{-iT^{(\rm{JC,Kerr})}H^{(\rm{JC,Kerr})}}\ket{\psi^{\rm (JC,Kerr)}_0}$, where $T^{(\rm{JC,Kerr})}$ is the physical evolution time for each non-linearity type. We define $\tilde{g}=T^{(\rm JC)}g$ and $\tilde{K}=T^{(\rm Kerr)}K$ as the adimensional parameters accounting for the interaction time.
    We label this strategy the \textit{continuous} approach.
    
    \item Dividing the total evolution time in discrete time steps and applying the non-linearities in quenches, alternating them with linear photon tunneling Hamiltonians $H^{(\rm t)} = J \left(a^\dagger_{2}a_{1} + a^\dagger_{1}a_{2}\right)$, and, in the JC case, also with free evolution Hamiltonians $H^{(\rm e)} = \Delta \sum^{2}_{i=1}\sigma^\dagger_i \sigma_i$ to account for potential phase differences between the emitter and photonic modes. The resulting state after the quenched evolution can be written as $\ket{\psi^{\rm (JC,Kerr)}_{\rm P}}=U^{\rm (JC,Kerr)}_{\rm P}\ket{\psi_0}$, where the unitary $U^{\rm (JC,Kerr)}_{\rm P}$ depends on the non-linearity applied. For the JC non-linearity, it reads:
    \begin{align}
        U^{\rm(JC)}_{\rm P} = \prod^d_{j=1} U^{(\rm JC)}_{j}U^{(\rm e)}_{j}U^{(\rm t)}_{j},
    \label{eq:Unitary JC}
    \end{align}
    where $d$ is the number of layers or quenches, $U^{(\rm t)}_{j}=e^{-iT^{\rm (t)}_j H^{(\rm t)}}$, $U^{(\rm e)}_{j}=e^{-iT^{\rm (e)}_j H^{(\rm e)}}$, and $U^{(\rm JC)}_{j}=e^{-iT^{\rm (JC)}_j H^{(\rm JC)}}$. For the Kerr-non linearity we build the unitary as:
    \begin{align}
        U^{\rm (Kerr)}_{\rm P} = \prod^d_{j=1} U^{\rm (Kerr)}_{j}U^{\rm (t)}_{j}.
    \label{eq:Unitary_Kerr}
    \end{align}
    where $U^{\rm (Kerr)}_{j}=e^{-iT^{\rm (Kerr)}_j H^{\rm (Kerr)}}$. 

    The key idea of this \emph{programmable} approach is that the unitaries $U^{\rm (JC,Kerr)}_{\rm P}(\boldsymbol{\pi})$ can be considered as \emph{parametrized quantum circuits} (PQCs), i.e., \emph{variational ansätze}. We refer to this step of the estimation process as the \emph{preparation} PQC. The two types of ansätze (labeled JC and Kerr) considered in this manuscript are sketched in Fig.~\ref{fig:sketch}(b,c). The parameters of the PQCs $\boldsymbol{\pi}$ can be optimized to minimize a given cost function. By choosing $C_{\rm P}(\boldsymbol{\pi})= -\mathcal{F}_{\rm Q}(\boldsymbol{\pi})$ as the cost-function we can find the optimal parameters $\boldsymbol{\pi}_{\rm opt}$ such that the resulting state after the unitary $\ket{\psi^{\rm (JC,Kerr)}_{\rm P}}=U^{\rm (JC,Kerr)}_{\rm P}(\boldsymbol{\pi}_{\rm opt})\ket{\psi_0}$ has the maximum potential metrological advantage. In the case of the Kerr ansatz, we use $\{\tilde{J}_j = J_j T^{\rm (t)}_{j}\}$ and $\{\tilde{K}_j = K_j T^{\rm (Kerr)}_{j}\}$ as variational parameters. For the JC ansatz we employ $\{\tilde{J}_j = J_j T^{\rm (t)}_{j}\}$, $\{\tilde{\Delta}_j = \Delta_j T^{\rm (e)}_{j}\}$, and $\{\tilde{g}_j = g_j T^{\rm (JC)}_{j}\}$ as variational parameters.

\end{itemize}

\subsection{Phase encoding} 
\label{sec:Phase encoding}

The phase encoding takes place after the preparation phase. Here, we use the standard strategy of photonic quantum metrology, i.e., a phase difference between the two arms of the MZI, see Fig.~\ref{fig:sketch}(a). This is described by the unitary $U^{\rm (MZ)}(\varphi)=U^{\rm (BS)}U^{\rm (PD)}(\varphi)U^{\rm (BS)}$, where $U^{\rm (BS)}=e^{-i\frac{\pi}{4}(a^\dagger_2 a_1 + a^\dagger_1 a_2)}$ is the unitary of a symmetric beam splitter and $U^{\rm (PD)}(\varphi)=e^{-i\frac{\varphi}{2}(a^\dagger_2 a_2 - a^\dagger_1 a_1)}$ encodes the phase difference between the two photonic modes. The output state after the phase encoding step is $\ket{\psi^{\rm (JC,Kerr)}_{\rm E}(\varphi)}=U^{\rm (MZ)}(\varphi)\ket{\psi^{\rm (JC,Kerr)}_{\rm P}}$.

\subsection{Measurement} 
\label{sec:Measurement}

Finally, to estimate the unknown parameter one needs to measure the state coming from the encoding phase, i.e.,  $\ket{\psi^{\rm (JC,Kerr)}_{\rm E}(\varphi)}$. Here, we will consider different strategies for the continuous and programmable approaches:
\begin{itemize}
    \item In the continuous approach, the output state of the encoding stage $\ket{\psi^{\rm (JC,Kerr)}_{\rm E}(\varphi)}$ is directly employed to assess the metrological power of a specific measurement. This is quantified by the CFI $\mathcal{F}_{\rm C}(\varphi)$, that requires calculating the probability distributions appearing in Eq.~\eqref{eq:CFI} as $P(\lambda|\varphi)=|\braket{\lambda|\psi^{\rm (JC,Kerr)}_{\rm E}(\varphi)}|^2$, which depend on the type of measurement employed. 
  
    \item For the programmable approach, we consider that the state after phase encoding $\ket{\psi^{\rm (JC,Kerr)}_{\rm E}(\varphi)}$ undergoes a second PQC, which we label \textit{pre-measurement} PQC to distinguish it from the preparation PQC. Each layer of the pre-measurement PQC is formed by the same gates as a layer of the preparation PQC, as sketched in Fig.~\ref{fig:sketch}(b,c). However, the pre-measurement PQC is described by a unitary $U^{\rm (JC,Kerr)}_{\rm M}(\boldsymbol{\mu})$, whose parameters ($\boldsymbol{\mu}$) are optimized to maximize the CFI for a given measurement. In the programmable approach, the optimizer computes the CFI from the probabilities $P(\lambda|\varphi)=|\braket{\lambda|\psi^{\rm (JC,Kerr)}_{\rm M}(\varphi)}|^2$ calculated with the state after the pre-measurement PQC, labeled as $\ket{\psi^{\rm (JC,Kerr)}_{\rm M}(\varphi)}=U^{\rm (JC,Kerr)}_{\rm M}(\boldsymbol{\mu})\ket{\psi^{\rm (JC,Kerr)}_{\rm E}(\varphi)}$.  
\end{itemize}

In this paper we consider two measurement types: photon counting and homodyne detection. Furthermore, for the JC non-linearity we assume that we also have access to the state of the quantum emitters coupled to the photonic modes. For the two approaches, one needs to define the correct positive operator-valued measure (POVM) to calculate the probability distributions $P(\lambda|\varphi)$ of the CFI~\eqref{eq:CFI}. For photon-counting:
\begin{itemize}
    \item The POVM of the Kerr ansatz uses only the eigenstates $\{\ket{N_1,N_2}\}$ of the number operators $n_i=a^\dagger_i a_i$ of the two photonic modes. From this, one obtains the probabilities $P(N_1,N_2|\varphi)=|\braket{N_1,N_2|\psi^{\rm (Kerr)}_{\rm E}(\varphi)}|^2$.

    \item For the JC ansatz, one needs to include as well in the POVM the eigenstates $z_i$ of the $\sigma_{\rm z,i}$ Pauli operators of the emitters, such that the final probabilities read $P(z_1,z_2,N_1,N_2|\varphi)=|\braket{z_1,z_2,N_1,N_2|\psi^{\rm (JC)}_{\rm E}(\varphi)}|^2$.
\end{itemize}

    In the case of homodyne detection:
    \begin{itemize}
        \item For the Kerr ansatz, one has to calculate the probabilities $P(x^{ (\theta)}_1,x^{ (\theta)}_2|\varphi)=|\braket{x^{ (\theta)}_1,x^{ (\theta)}_2|\psi^{\rm (Kerr)}_{\rm E}(\varphi)}|^2$ of obtaining a measurement outcome $x^{ (\theta)}_1$, $x^{ (\theta)}_2$ when one measures the generalized quadrature $X_i(\theta)=(e^{-i\theta}a^\dagger_i+e^{i\theta}a_i)/\sqrt{2}$ of each mode $i=1,2$, where $\theta$ is the quadrature angle.

        \item For the JC ansatz, the probabilities are instead $P(z_1,z_2,x^{ (\theta)}_1,x^{ (\theta)}_2|\varphi)=|\braket{z_1,z_2,x^{ (\theta)}_1,x^{ (\theta)}_2|\psi^{\rm (JC)}_{\rm E}(\varphi)}|^2$, since the POVM in this case is formed by the eigenstates of the $\sigma_{\rm z,1}\otimes\sigma_{\rm z,2}\otimes X_1(\theta)\otimes X_2(\theta)$ operator, i.e., $\{\ket{z_1,z_2,x^{(\theta)}_1,x^{(\theta)}_2}\}$.
        
    \end{itemize}
    
  Finally, let us note an important difference between the continuous and programmable approaches. In the former, we choose the optimal  $\theta$ for which the CFI is maximized [see Sec.~SM2 of the SM]. On the contrary, in the latter we fix $\theta=0$ since the pre-measurement PQC already allows to optimize the CFI [see Sec.~SM3 of the SM]. In Sec.~SM4 of the SM we also consider an alternative strategy based on fixing $U^{\rm (JC,Kerr)}_{\rm M}=\mathbb{1}$ and optimizing the quadrature angle, finding similar results to those presented in the manuscript.

\begin{figure*}[tb]
    \centering
    \includegraphics[width=\linewidth]{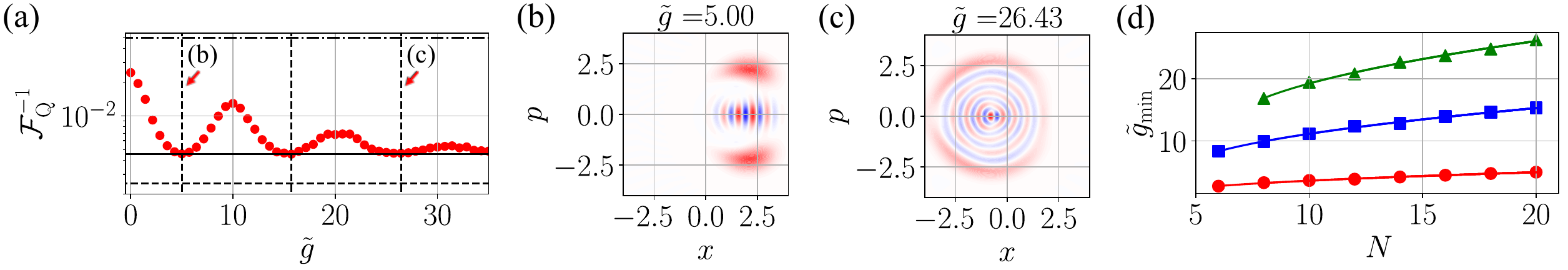}
    \caption{ (a) Inverse QFI $\mathcal{F}^{-1}_{\rm Q}$ (red dots) for continously generated states using the JC non-linearity and an initial state with mean-photon number $N=20$. Horizontal lines: dashed, solid, and dashed-dotted lines correspond to the HL, the TFS results, and the SQL, respectively. The vertical lines signal the position of the minima. (b,c) Wigner functions of the displaced cat state and the displaced Fock state that appear at the values of $\tilde{g}$ signalized by the first and third vertical lines in panel (a). (d) Interaction time $\tilde{g}_{\rm min}$ at which $\mathcal{F}^{-1}_{\rm Q}$ reaches its first (red circles), second (blue squares), and third (green triangles) minimum as a function of $N$. Circles are numerical results, while solid lines correspond to square-root fits. }
    \label{fig:dynamics JC}
\end{figure*}

\section{Probe state generation in the continuous approach} 
\label{sec:continuous}

In this Section we study the probe states generated with the continuous evolution, focusing on the maximal potential metrological advantage that they can provide (given by the QFI). Sec.~\ref{sec:continuous JC} analyzes the results for the JC non-linearity, while Sec.~\ref{sec:continuous Kerr} focuses on the Kerr interaction.

\subsection{JC non-linearity}
\label{sec:continuous JC}

We start with the states that can be generated using the JC non-linear Hamiltonian~\eqref{eq:interaction JC}. Here, the input state $\ket{\psi^{\rm (JC)}_0}=\ket{g,g,\alpha,\alpha}$ undergoes a unitary evolution given by $\ket{\psi^{\rm (JC)}_{\rm P}(\tilde{g})}=e^{-iT^{(\rm{JC})}H^{(\rm{JC})}(g)}
\ket{\psi^{\rm (JC)}_0}$, where $\tilde{g}=T^{\rm (JC)}g$ is the adimensional interaction time. To assess the metrological power of $\ket{\psi^{\rm (JC)}_{\rm P}(\tilde{g})}$, we calculated its QFI for phase estimation in an MZI. In Fig.~\ref{fig:dynamics JC}(a) we show the inverse of the QFI, $\mathcal{F}^{-1}_{\rm Q}$, for two initial coherent states featuring a total mean number of photons $N=20$. The QFI displays an oscillatory behavior: $\mathcal{F}^{-1}_{\rm Q}$ first decreases down to the estimation error set by TFS with $10$ photons in each mode, i.e., $\ket{10}\otimes\ket{10}$. Then, $\mathcal{F}^{-1}_{\rm Q}$ increases again, but it does not recover its initial value due to the dephasing of the different photon-number components of the coherent states introduced by the non-linearity. Similar dumped oscillations are known to arise in the probability of measuring a single quantum emitter in its ground or excited state~\cite{Shore1993}.

We now study the nature of the states produced by the protocol, particularly the ones minimizing $\mathcal{F}^{-1}_{\rm Q}$. To do it, in Fig.~\ref{fig:dynamics JC}(b,c) we plot the Wigner quasiprobability distribution of a single photonic mode (obtained by tracing the other photonic mode as well as the two quantum emitters, i.e., $\tr_{ee1}\{\ket{\psi^{\rm (JC)}_{\rm P}(\tilde{g})}\bra{\psi^{\rm (JC)}_{\rm P}(\tilde{g})}\}$) at values of the adimensional interaction time $\tilde{g}\simeq 5$ and $\tilde{g}\simeq 26$, corresponding to the first and third minima of $\mathcal{F}^{-1}_{\rm Q}$. The Wigner distribution at such values of $\tilde{g}$ corresponds respectively to a displaced cat state along the $x$ quadrature, and a displaced Fock state, also along the $x$ quadrature as already shown in a previous work~\cite{Uria2020}. The second minimum of $\mathcal{F}^{-1}_{\rm Q}$, appearing at $\tilde{g}\simeq 16$, corresponds to a state displaying phase-space oscillations but that cannot be associated with any well-known type.

As a next step, we calculated the dependence with $N$ of the values of $\tilde{g}$ at which the minima of $\mathcal{F}^{-1}_{\rm Q}$ take place, which we label $\tilde{g}_{\rm min}$. Fig.~\ref{fig:dynamics JC}(d) shows that, for the first three minima, $\tilde{g}_{\rm min}$ follows a $\sqrt{N}$ dependence. This is illustrated with a square root fit of the type $\tilde{g}_{\rm min}=\alpha\sqrt{N+\beta}+\gamma$ for each minimum, which reproduces the behavior of the data points with great precision. Actually, such a scaling with $N$ was already observed for the displaced Fock state in Ref.\cite{Uria2020}. However, an important observation of our analysis is that metrologically useful of states with the same potential that displaced Fock states can be obtained at shorter interaction times [red line vs green line in Fig.~\ref{fig:dynamics JC}(d)].

\subsection{Kerr non-linearities}
\label{sec:continuous Kerr}

\begin{figure*}[tb]
    \centering
    \includegraphics[width=\linewidth]{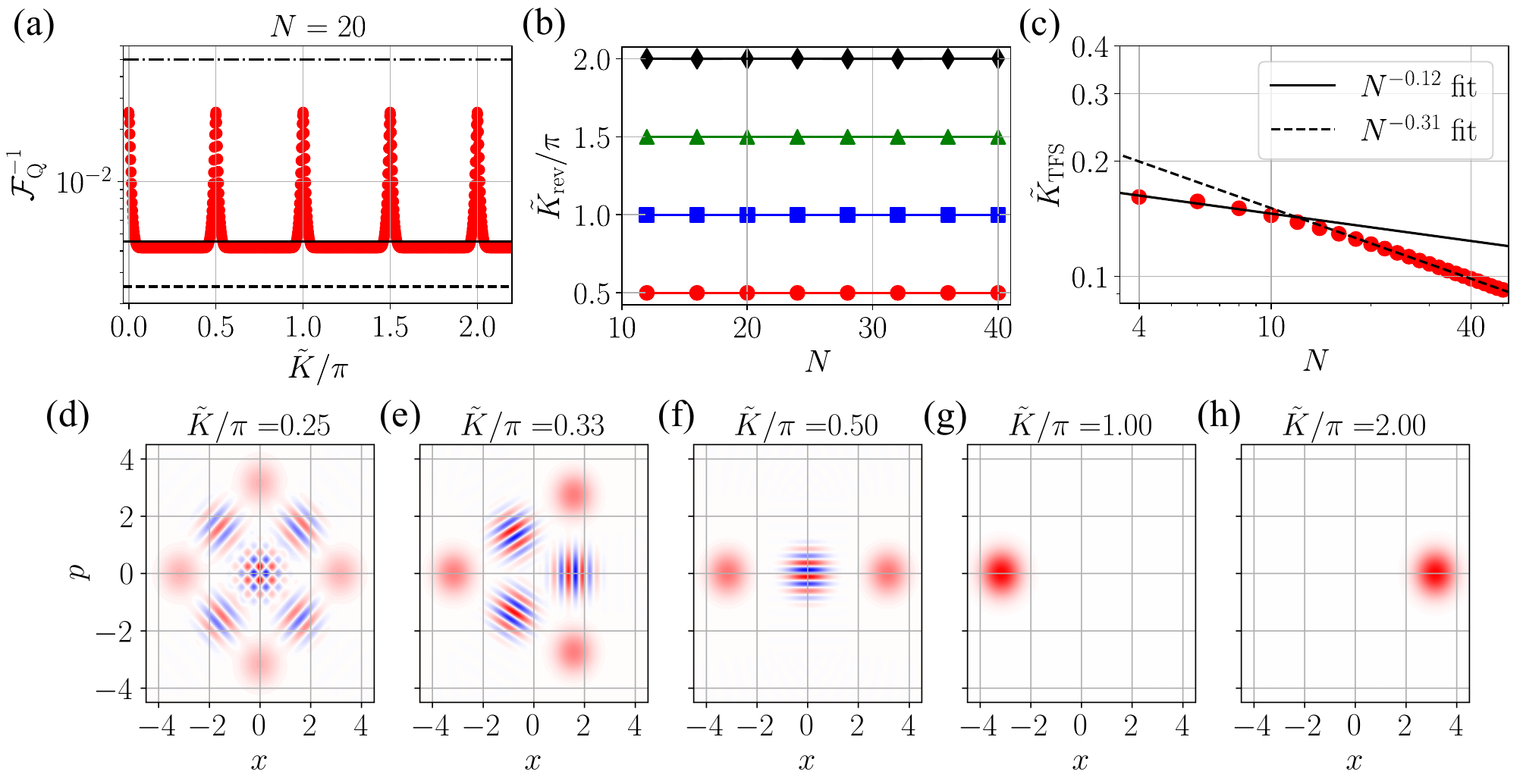}
    \caption{(a) Inverse QFI $\mathcal{F}^{-1}_{\rm Q}$ (red dots) for continuously generated states using the Kerr non-linearity and an initial state with mean-photon number $N=20$. (b) Interaction time $\tilde{K}_{\rm rev}$ of the first (red circles), second (blue squares), third (green triangles), and fourth (black rhombus) revivals of $\mathcal{F}^{-1}_{\rm Q}$ as a function of $N$. Solid lines are guides for the eye. (c) Interaction time $\tilde{K}_{\rm TFS}$ necessary for the QFI of the generated probe states to equal that of TFS as a function of $N$ (red circles). The lines correspond to linear fits. (d-h) Wigner function of a single photonic mode for $N=20$ at several values of the interaction time $\tilde{K}$.}
    \label{fig:dynamics nk}
\end{figure*}

We now repeat the previous analysis for an initial state $\ket{\psi^{\rm (Kerr)}_0}$ evolving under the Kerr Hamiltonian  $H^{\rm (Kerr)}$ given by Eq.~\eqref{eq:interaction Kerr}. The output state of the probe preparation stage reads
\begin{align}
    \ket{\psi^{\rm (Kerr)}_{\rm P}(\tilde{K})} &= e^{-iT^{\rm (Kerr)}H^{\rm (Kerr)}(K)}\ket{\psi^{\rm (Kerr)}_0},  
\label{eq:probe state Kerr}
\end{align}
with $\tilde{K}=T^{\rm (Kerr)}K$ being the adimensional interaction time. The output state $\ket{\psi^{\rm (Kerr)}_{\rm P}(\tilde{K})}$ is fed to the MZI to compute the QFI. The results for $\mathcal{F}^{-1}_{\rm Q}$ as a function of the interaction time $\tilde{K}$ are shown in Fig.~\ref{fig:dynamics nk}(a) for a total mean photon number $N=20$. In this case, the inverse of the QFI features a similar behavior to that observed for the JC interaction: starting from a large value corresponding to the initial coherent states, $\mathcal{F}^{-1}_{\rm Q}$ first decreases slightly below the error bound obtained by TFS $\ket{N/2}\otimes\ket{N/2}$, and then increases again. However, differently from the JC interaction, the Kerr non-linearity allows the QFI to recover its initial value. Furthermore, notice that such revivals appear at integer multiples of $\tilde{K}=\pi/2$ independently of $N$, as it is shown in Fig.~\ref{fig:dynamics nk}(b) for the first four revivals. The reason behind this behavior is that the $n^2$ non-linearity of the Kerr Hamiltonian allows for a rephasing of the photon-number components at such times resulting in either coherent states or $2$-component cat states [see panels (f-h) for the Wigner function of a single mode at the first, second, and fourth revivals. At the third revival a $2$-component cat state is generated]. However, none of these state provides the best potential metrological advantage.

The most promising states are actually generated in the plateaus between the QFI revivals of Fig.~\ref{fig:dynamics nk}(a). Here the
generated probe states $\ket{\psi^{\rm (Kerr)}_{\rm P}(\tilde{K})}$ evolve between different multicomponent cat states. In particular, at values of the non-linearity time $\tilde{K}=\pi(\ell + 1/q)$, where $\ell=0,1,2,...$ and $q=...-2,-1,1,2,...$, a $q$-component cat state is produced~\cite{Haroche2006exploring,Kirchmair2013}. For $q\geq 3$ the $q$-component cat states feature a QFI comparable to that of TFS. This is shown explicitly in Fig.~\ref{fig:dynamics nk}(d,e), where we plot the Wigner distribution for $\ell=0$ and $q=3,4$. For such values of $\tilde{K}$, Fig.~\ref{fig:dynamics nk}(a) shows that these states feature an $\mathcal{F}^{-1}_{\rm Q}$ slightly smaller than that of TFS with the same $N$.

\begin{figure}[tb]
    \centering
    \includegraphics[width=\linewidth]{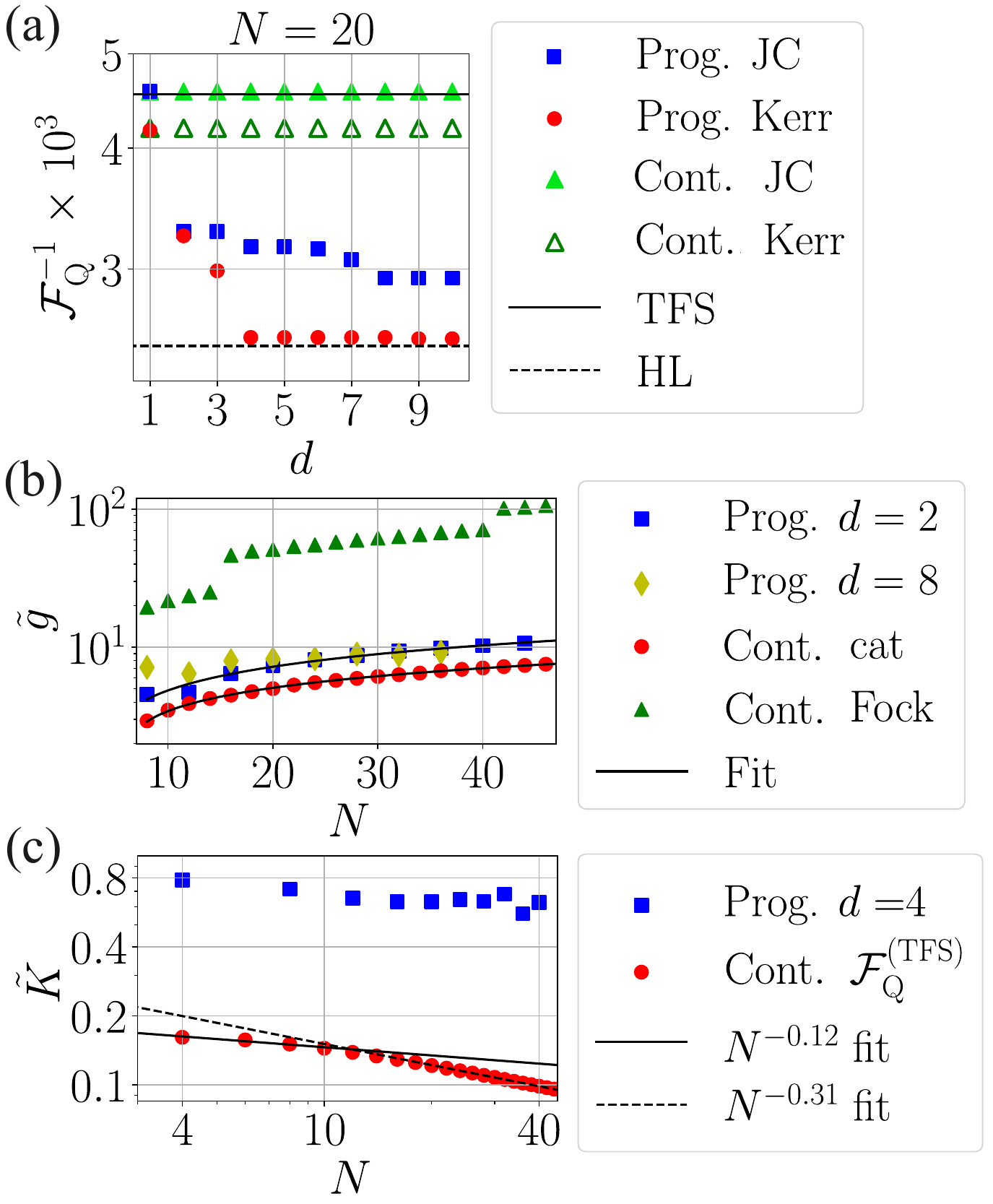}
    \caption{
    (a) Inverse QFI $\mathcal{F}^{-1}_{\rm Q}$  as a function of the number of layers $d$ in the JC (blue squares) and Kerr (red circles) ansätze of the programmable approach for a mean-photon number $N=20$. The first minimum of $\mathcal{F}^{-1}_{\rm Q}$ obtained by the continuous approach employing a JC (Kerr) interaction is signaled by the solid (empty) green triangles. Although they do not depend on $d$, we plot them as points instead of lines for better figure clarity. Black dashed (solid) lines signal the HL (TFS results). 
    (b) JC non-linearity: interaction time $\tilde{g}$ needed to produce states with the smallest value of $\mathcal{F}^{-1}_{\rm Q}$ as a function of $N$ for each strategy. Blue squares (yellow rhombus): sum of the absolute value of all interaction optimal parameters $\tilde{g}_{\rm T}$ obtained by the programmable approach with $d=2$ ($d=8$) Green triangles: displaced Fock states obtained with the continuous approach (data taken from Ref.~\cite{Uria2020}). Red circles:  displaced cat states obtained with the continuous approach. Lines are square root fits.
    (c) Kerr non-linearity: interaction time $\tilde{K}$ needed to produce states with the smallest value of $\mathcal{F}^{-1}_{\rm Q}$ as a function of $N$ for each strategy. Blue squares: sum of the absolute value of all interaction optimal parameters $\tilde{K}_{\rm T}$ obtained by the programmable approach with $d=4$. Red circles: interaction time $\tilde{K}_{\rm TFS}$ necessary to produce states with the same QFI as TFS in the continuous approach. Lines are linear fits.
    }
    \label{fig:state preparation benchmark}
\end{figure}

In analogy with what we do with the JC non-linearity, we study the interaction time $\tilde{K}_{\rm TFS}$ that the Kerr non-linearity needs to transform the initial state $\ket{\psi^{\rm (Kerr)}_0}=\ket{\alpha,\alpha}$ with $|\alpha|^2=N/2$ into a state featuring the same value of the QFI as the TFS $\ket{N/2,N/2}$. This is shown in Fig.~\ref{fig:dynamics nk}(c). Differently from the JC model, $\tilde{K}_{\rm TFS}$ decreases with $N$, meaning that larger probe states need smaller interaction times to attain the metrological advantage of TFS. In the case of the Kerr interaction, the dependence of $\tilde{K}_{\rm TFS}$ with $N$ is fit by two power laws: for $N\lesssim 10$ we have $\mu=-0.12$ and for $N\gtrsim 10$ we have $\mu=-0.31$. Our intuition is that the quadratic dependence of the Kerr non-linearity on the photon number enables a faster non-trivial state preparation, although we cannot explain analytically the asymptotic behavior displayed in Fig.~\ref{fig:dynamics nk}(c).

Summing up, for both the JC and the Kerr interaction the continuous application of the non-linearity allows one to obtain states with a QFI similar to TFS, but it does not allow to saturate the HL. In the next section, we show that the programmable approach offers a path to reach the HL using the same type of non-linearities and requiring comparable interaction times.

\section{Probe state generation in the programmable approach}
\label{sec:Programmable}

In this Section we apply the programmable approach described in Sec.~\ref{sec:Preparation} and compare it with the performance of the continuous approach.

We start by benchmarking the values of QFI obtained with the two approaches to quantify the metrological advantage of the probe states generated in each case. In Fig.~\ref{fig:state preparation benchmark}(a) we show the inverse of the QFI, $\mathcal{F}^{-1}_{\rm Q}$, for an initial state with a fixed mean-photon number $N=20$ as a function of the number of layers $d$ employed by the programmable approach. Results for the JC (Kerr) ansatz are shown as blue squares (red circles). For reference, we also plot in black solid and dashed lines the QFI obtained by TFS and the HL, respectively, and the best QFI obtained in the continuous approach for the JC (Kerr) non-linearity in filled (empty) triangles. The programmable approach is able to improve the results obtained by continuous evolution as we increase the number of layers of the circuit, even almost saturating the HL with only four layers in the case of the Kerr ansatz.

While this improvement with the number of layers (and thus with the number of operations) was already observed in our previous work~\cite{MunozDeLasHeras_2024}, whether it comes at the price of larger total interaction times when considering all layers was still a relevant open question. The total interaction time can be calculated by summing the interaction times through all the circuit, i.e., $\tilde{g}_{\rm T} = \sum^d_{i=1} |\tilde{g}_i|$ and $\tilde{K}_{\rm T} = \sum^d_{i=1} |\tilde{K}_i|$, for the JC and Kerr ansätze, respectively. It is important to have a total interaction time as small as possible because errors produced, e.g., by photon loss, depend on the total interaction time and not on the number of layers.

In Figs.~\ref{fig:state preparation benchmark}(b,c), we address this question by showing the total interaction time required by the programmable and continuous approaches to reach their maximum QFI as a function of the mean photon number of the initial state. In Fig.~\ref{fig:state preparation benchmark}(b) we plot $\tilde{g}_{\rm T}$ as a function of $N$ for JC ansätze with $d=2$ ($d=8$) in blue squares (yellow rhombus). The values of $\tilde{g}_{\rm T}$ for the same $N$ are very similar irrespectively of the number of layers $d$ of the ansatz, which means that the optimization can improve the QFI without increasing the total interaction time. A $\sqrt{N}$-dependence similar to that observed in the continuous approach is found for $\tilde{g}_{\rm T}$ irrespectively of the number of layers, although it becomes less clear as one increases $d$. To shine more light on the comparison between the continuous and the programmable approaches, we plot the time that the former strategy requires to generate displaced Fock (green triangles, data obtained by Ref.~\cite{Uria2020}) and cat states (red circles). The programmable approach requires total interaction times that lie between those of the two state classes generated with the continuous strategy. In Fig.~\ref{fig:state preparation benchmark}(c) we perform the same analysis for the Kerr non-linearity ansatz, and we benchmark the results of the optimal preparation PQC with $d=4$ layers (in blue squares) against those obtained by continuous evolution (in red circles). Here we want to highlight two results: i) The programmable Kerr ansatz is able to saturate the HL at the expense of slightly larger interaction times than the continuous approach. ii) Like in the continuous strategy, the programmable interaction time also decreases with the photon number. This implies that Kerr-non linearities open a path to generate large photon states approaching the HL using the programmable approach. For the two ansätze the nature of the photonic states generated with $d\geq 2$ using the programmable strategy cannot be ascribed to any previously known state (see Sec.~SM5 of the SM for more details on their Wigner distribution).

\section{Measurements}
\label{sec:Measurements}

\begin{figure}[tb]
    \centering
    \includegraphics[width=\linewidth]{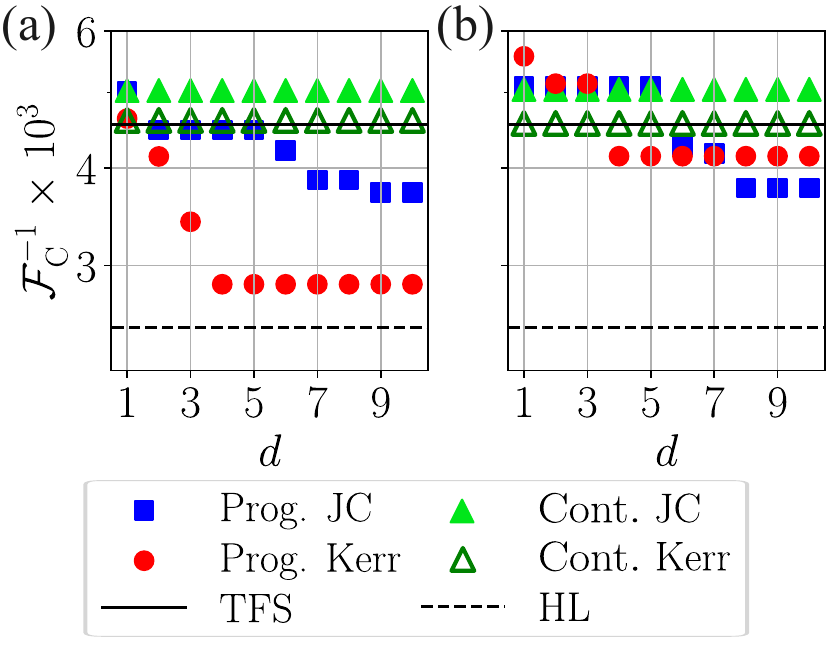}
    \caption{
   (a,b) Inverse CFI $\mathcal{F}^{-1}_{\rm C}$  as a function of the number of layers $d$ in the JC (blue squares) and Kerr (red circles) ansätze of the programmable approach for a mean-photon number $N=20$ for photon counting (a) and homodyne detection (b). The smallest values of $\mathcal{F}^{-1}_{\rm C}$ obtained by the continuous approach employing a JC (Kerr) interaction are signaled by the solid (empty) green triangles. Although they do not depend on $d$, we plot them as points instead of lines for better figure clarity. Black dashed (solid) lines signal the HL (TFS results). 
    }
    \label{fig:measurement}
\end{figure}

Let us finally focus on the last part of photonic quantum metrology protocols, that is, the measurement of the probe state to estimate the encoded parameter. In particular, we calculate the CFI for two different measurements, that are, photon counting and homodyne detection, as explained in Sec.~\ref{sec:Measurement}.

In Fig.~\ref{fig:measurement}(a) we show the CFI for photon-counting measurements corresponding to the best probe states generated by continuous JC and Kerr non-linear evolution [i.e., those that correspond to the first minima of $\mathcal{F}^{-1}_{\rm Q}$ in Figs.~\ref{fig:dynamics JC}(a) and~\ref{fig:dynamics nk}(a)] in filled and empty triangles, respectively, for an initial state with mean-photon number $N=20$. The Kerr evolution is able to obtain the CFI of the TFS, whereas the JC non-linearity performs slightly worse. However, the important part of this figure is the evolution of the CFI as a function of the number of layers $d$ for both the JC (blue squares) and Kerr (red circles) programmable ansätze. Like for the QFI, both ansätze are able to go below the precision attainable by TFS, although in this case they do not saturate the HL.

In Fig.~\ref{fig:measurement}(b) we carry a similar analysis for both the continuous and programmable approaches, but in this case by considering homodyne detection as the measurement protocol. The quadrature angle is chosen as explained in Section~\ref{sec:Measurement}. For the continuously generated states, both the JC and the Kerr non-linearities give values of $\mathcal{F}^{-1}_{\rm C}$ very similar to those obtained using photon counting, approaching the CFI of TFS in the former case and saturating it in the latter. In the programmable approach, the pre-measurement optimization allows to reach larger values of the CFI than those of TFS. However, for this strategy homodyne detection performs worse than photon counting.

\section{Conclusions \& Outlook}
\label{sec:Conclusions}

Summing up, we present a systematic study of the potential of programmable photonic non-linearities to improve the generation of metrologically-useful quantum states of light. By studying the quantum Fisher information of the generated states, we demonstrate that the programmable strategy reaches better precisions than continuous time evolution  using similar interaction times. When one employs the former strategy, while for the Jaynes-Cummings non-linearity the total interaction time grows with the number of photons, for the Kerr non-linearity the total interaction time decreases with increasing photon number, and the generated states approach the Heisenberg limit. Finally, we study the role of the measurement in the estimation process, finding that photon counting performs better than homodyne detection for the generated states, and we benchmark the improvement that adding a pre-measurement programmable quantum circuit produces. As an outlook, in future works we will explore other approaches for quantum state generation, including driven-dissipative settings~\cite{Kraus2008,Diehl2008,Verstraete2009,Barreiro2011,Lin2013} or by using other non-linearity types~\cite{FriskKockum2019,Chang2020,Agusti2020,Agusti2022}. Another possible direction is to harness the programmable photonic circuits in the context of state preparation for bosonic error-correcting codes~\cite{Devitt_2013,Cai2021}, such as GKP states~\cite{Gottesman2001,Vasconcelos2010,Shi_2019,Grimsmo2021,Hastrup2022}, which can also be useful for quantum metrology~\cite{gardner2024stochasticwaveformestimationfundamental}.

\begin{acknowledgements}
  The authors acknowledge support from the Proyecto Sin\'ergico CAM 2020 Y2020/TCS-6545 (NanoQuCo-CM), the CSIC Research Platform on Quantum Technologies PTI-001 and from Spanish projects PID2021-127968NB-I00 and TED2021-130552B-C22 funded by  MCIN/AEI/10.13039/501100011033/FEDER, UE and MCIN/AEI/10.13039/501100011033, respectively. AMH acknowledges support from Fundación General CSIC's ComFuturo program, which has received funding from the European Union's Horizon 2020 research and innovation program under the Marie Skłodowska-Curie grant agreement No. 101034263. AGT also acknowledges support from the QUANTERA project MOLAR with reference PCI2024-153449 and funded MICIU/AEI/10.13039/501100011033 and by the European Union. The authors also acknowledge Centro de Supercomputación de Galicia (CESGA) who provided access to the supercomputer FinisTerrae for performing numerical simulations. 
\end{acknowledgements}


\bibliography{cleaned_references}

\clearpage

\begin{widetext}

\begin{center}
\textbf{\large Supplementary Material: Improving quantum metrology protocols with programmable photonic circuits\\}
\end{center}
\setcounter{equation}{0}
\setcounter{figure}{0}
\setcounter{section}{0}
\makeatletter

\renewcommand{\thefigure}{SM\arabic{figure}}
\renewcommand{\thesection}{SM\arabic{section}}  
\renewcommand{\theequation}{SM\arabic{equation}}  

In this Supplementary Material, we provide more details on our manuscript entitled \emph{``Improving quantum metrology protocols with programmable photonic circuits''}. In Sec.~\ref{sec SM: numerics} we give more details on how our numerical simulations are carried. Sec.~\ref{sec SM: theta min determination} provides a detailed description of the process employed to find the optimal quadrature angles for homodyne detection using the continuous approach. In Sec.~\ref{sec SM: optimization theta} we compare the results obtained for homodyne detection in the programmable approach when one optimizes the quadrature angle or, alternatively, fixes its value. In Sec.~\ref{sec SM: measurement layers} we compute the value of the classical Fisher information (CFI) of the states produced by the programmable approach without including the pre-measurement optimization loop. Finally, in Sec.~\ref{sec SM: probe states programmable} we examine the Wigner quasi-probability distribution of the probe states generated with the programmable approach.

\section{Details on the numerical simulations}
\label{sec SM: numerics}

Here we provide more details on the numerical simulations performed to calculate the results shown in this manuscript. 

For both the continuous and the programmable approaches, we perform full state-vector simulations of the photonic quantum states. In practice, one needs to introduce a cutoff for the Hilbert space dimension. In our numerics, we set this cutoff to $2N$ for each mode, where $N$ is the total mean number of photons in the two optical modes. To simulate the phase estimation process, one needs to assume a certain value of the phase $\varphi$ that one aims at estimating. In our case, we set $\varphi=\pi/3$. To calculate the quantum Fisher information (QFI) according to Eq.~(1) of the Main Text, we take $\delta=10^{-2}$. We have made sure that for such a small value the results of the QFI are converged with respect to $\delta$. 

In the programmable approach, we employ the COBYLA method to perform the classical optimization of both the preparation and the pre-measurement parametrized quantum circuits (PQCs). This choice was made after benchmarking COBYLA against different optimization algorithms: BFGS, L-BFGS-B, and SLSQP. COBYLA was the one giving the best results within reasonable computation times. Each optimization starts from random initial parameters close to zero for both the preparation and measurement PQCs (such that, initially, the unitaries of the two PQCs are close to the identity matrix). For each value of the mean photon number $N$, we initialize both the preparation and measurement PQCs with a single layer. Once the optimization of the two PQCs finishes (after 1000 iterations or a convergence tolerance of $10^{-10}$), we add another layer to both PQCs. The new initial parameters for the first layer are the optimal ones, while the initial parameters of the new layer are set to zero. We repeat this process for growing values of $d$ up to $d=10$. Once this optimization series finishes, we change the seed for the initial parameters of the single-layer PQCs and we repeat the process once again from $d=1$ to $d=10$. Results are gathered for $60$ different sets of initial parameters in the case of the JC ansatz and $40$ sets for the Kerr ansatz. The data shown in the Main Text for each value of $N$ and $d$ correspond to the results featuring the largest values of QFI and CFI within the whole pool of optimization runs.

\section{Optimization of the quadrature angles with the continuous approach}
\label{sec SM: theta min determination}

In this Section, we determine the optimal angles $\theta_{\rm min}$ for the generalized quadratures $X_i(\theta)=(e^{-i\theta}a^\dagger_i+e^{i\theta}a_i)/\sqrt{2}$ to be measured in homodyne detection using the continuous approach, where $i=1,2$ is the index of each photonic mode. This process is carried out independently for each interaction type. 

We start by fixing the adimensional interaction time at values producing a minimum of the inverse of the CFI $\mathcal{F}^{-1}_{\rm C}$. For a mean-photon number $N=20$, this corresponds to $\tilde{g}\simeq 5$ and $\tilde{K}=\pi/4$ for the Jaynes-Cummings (JC) and Kerr nonlinearities, respectively, see Fig.~2(a) and Fig.~3(a) of the Main Text. For those values of the interaction strength, we then calculate $\mathcal{F}^{-1}_{\rm C}$ as a function of $\theta$, following the procedure described in Sec.~II of the Main Text. The results are shown in panels (a,b) of Fig.~\ref{fig SM: theta min determination} for the JC and the Kerr interaction, respectively. In the two cases, $\mathcal{F}^{-1}_{\rm C}$ displays an oscillatory behavior as a function of $\theta$. For the JC interaction, there are two points located at $\theta_{\rm min}\simeq 2\pi/3$ and $\theta_{\rm min}\simeq 5\pi/3$ where $\mathcal{F}^{-1}_{\rm C}$ takes its global minimum value. Without loss of generality, we use the first one (i.e., $\theta_{\rm min}\simeq 2\pi/3$) to calculate the optimal generalized quadrature $X(\theta_{\rm min})$. In the case of the Kerr nonlinearity, there are several values of $\theta$ for which $\mathcal{F}^{-1}_{\rm C}$ reaches its global minimum. Again, without loss of generality, we use the first one, located at $\theta_{\rm min}=0.17\pi$, to calculate $X(\theta_{\rm min})$.

Finally, in Fig.~\ref{fig SM: theta min determination}(c) we plot the values of $\theta_{\rm min}$ where $\mathcal{F}^{-1}_{\rm C}$ reaches its first global minimum as a function of $N$. As shown in the figure, their position is independent of $N$ for the two interaction types. Thus, we use the same values of $\theta_{\rm min}$ to compute the generalized quadratures in homodyne detection regardless of $N$.

\section{Optimization of the quadrature angle with the programmable approach}
\label{sec SM: optimization theta}

\begin{figure*}[tb]
    \centering
    \includegraphics[width=\linewidth]{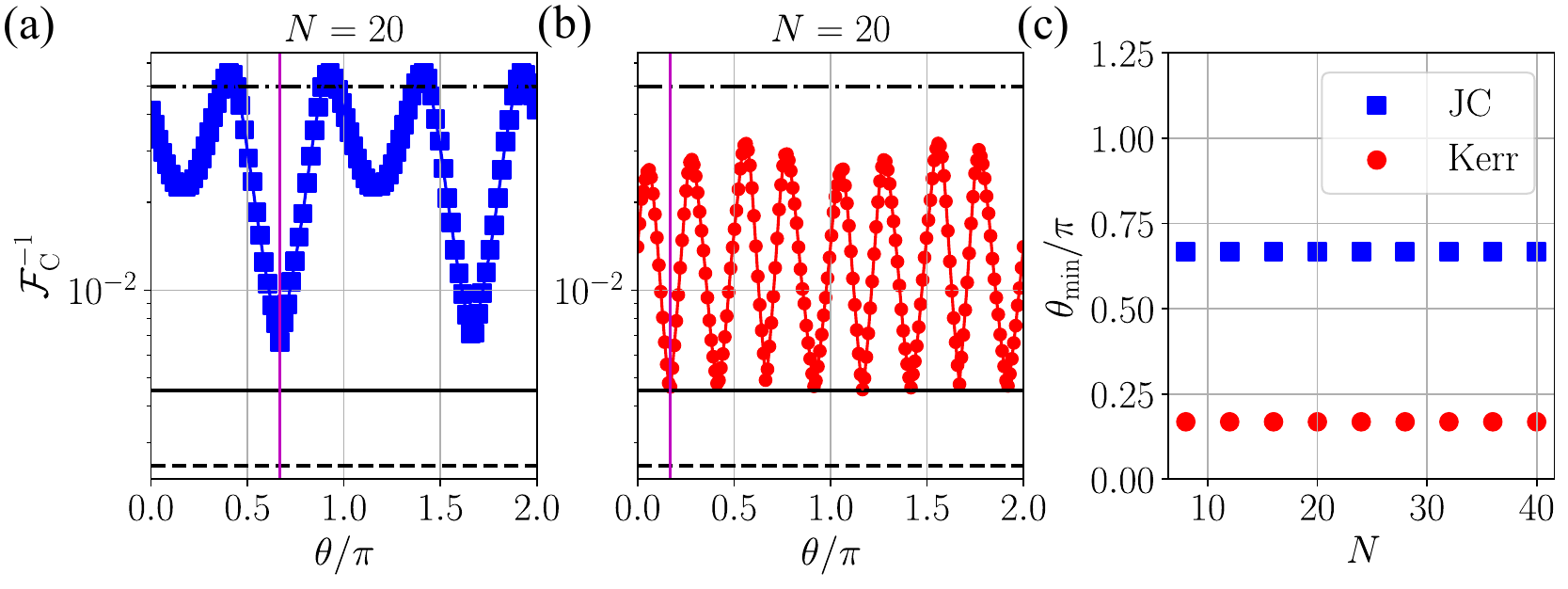}
    \caption{(a,b) Inverse CFI $\mathcal{F}^{-1}_{\rm C}$ as a function of the generalized quadrature angle $\theta$ for homodyne detection. The mean-photon number of the initial state is fixed at $N=20$. The horizontal dashed, solid, and dashed-dotted black lines represent the Heisenberg limit, the twin-Fock states results, and the standard quantum limit, respectively. Squares and circles are the data obtained in the continuous approach using the  JC [panel (a)] and the Kerr interaction [panel (b)]. The adimensional interaction time is fixed at $\tilde{g}\simeq 5$ and $\tilde{K}=\pi/4$ for each interaction type. (c) Location $\theta_{\rm min}$ of the minima of $\mathcal{F}^{-1}_{\rm C}$ signalized by the vertical magenta lines in panels (a) and (b) as a function of $N$. Blue squares (red circles) are the results for the JC (Kerr) interaction.}
    \label{fig SM: theta min determination}
\end{figure*}

\begin{figure*}[tb]
    \centering
    \includegraphics[width=0.9\linewidth]{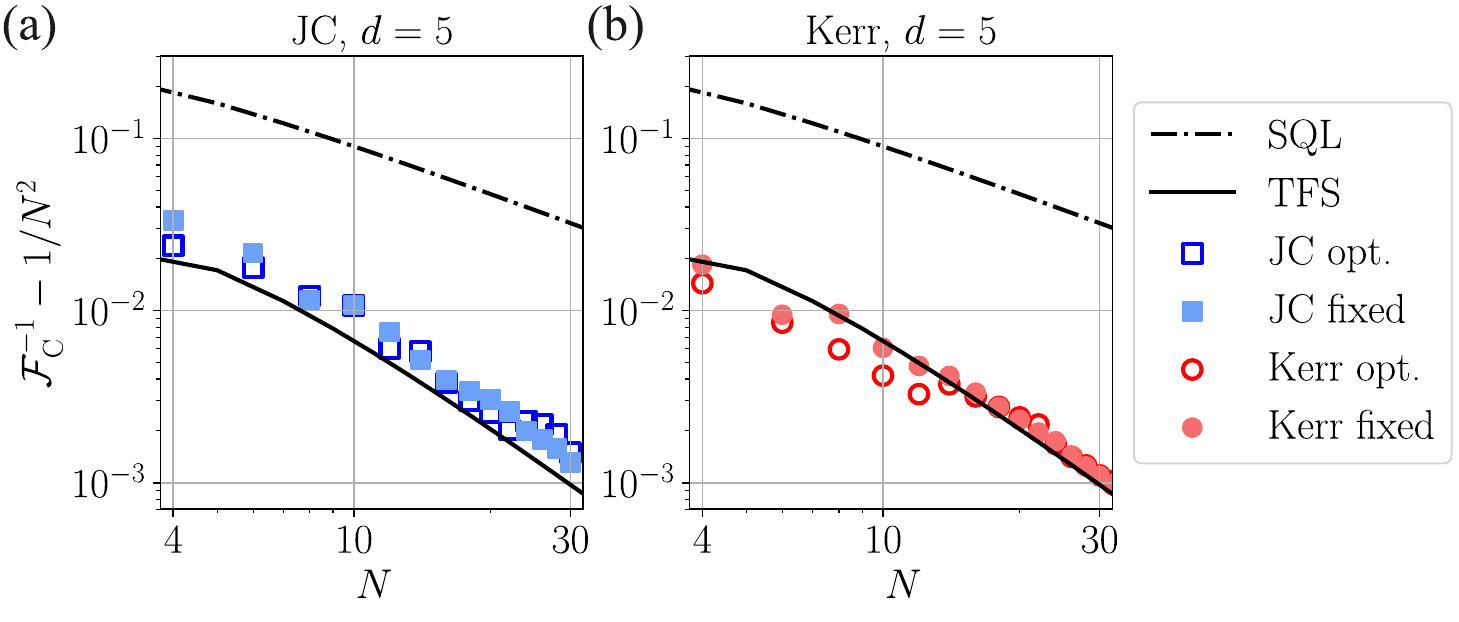}
    \caption{Inverse CFI $\mathcal{F}^{-1}_{\rm C}$ as a function of the mean-photon number $N$ obtained in homodyne detection with the programmable approach using the JC [panel (a)] and the Kerr [panel (b)] ansätze. The number of layers of the preparation and pre-measurement PQCs is fixed to $d=5$. Solid markers correspond to the results obtained by fixing the quadrature angle to $\theta=0$, while void markers correspond to the results produced by optimizing the quadrature angle $\theta$ as another variational parameter. In the two cases, the pre-measurement PQC is optimized. In both panels, we plot the difference between the data and the Heisenberg limit, $1/N^2$. Solid (dashed-dotted) lines represent the twin-Fock states results (standard quantum limit).}
    \label{fig SM: optimization theta programmable}
\end{figure*}

In this Section, we study the CFI obtained in homodyne detection using the programmable approach. We compare two strategies: the first one consists on optimizing the angle $\theta$ of the generalized quadratures $X_i(\theta)=(e^{-i\theta}a^\dagger_i+e^{i\theta}a_i)/\sqrt{2}$ (where $i=1,2$ is the index of each photonic mode) as an additional variational parameter of the pre-measurement PQC. The second strategy only optimizes the pre-measurement PQC, while employing a fixed $\theta=0$. 

The results comparing the two approaches are shown in Fig.~\ref{fig SM: optimization theta programmable}, where we plot the inverse of the CFI $\mathcal{F}^{-1}_{\rm C}$ as a function of the mean-photon number $N$ for the JC and the Kerr ansätze with $d=5$ layers. For both ansätze, the two optimization strategies produce very similar results. This is because the measurement PQC can effectively rotate the Wigner function of the quantum state in phase space. Thus, it has a similar effect than changing the angle $\theta$ of the measured quadrature. This demonstrates that, in the programmable approach, it is not necessary to optimize $\theta$ for homodyne detection. This is why in the Main Text we consider results with a fixed $\theta=0$.

\section{Effect of the pre-measurement parametrized quantum circuit}
\label{sec SM: measurement layers}

\begin{figure*}[tb]
    \centering
    \includegraphics[width=\linewidth]{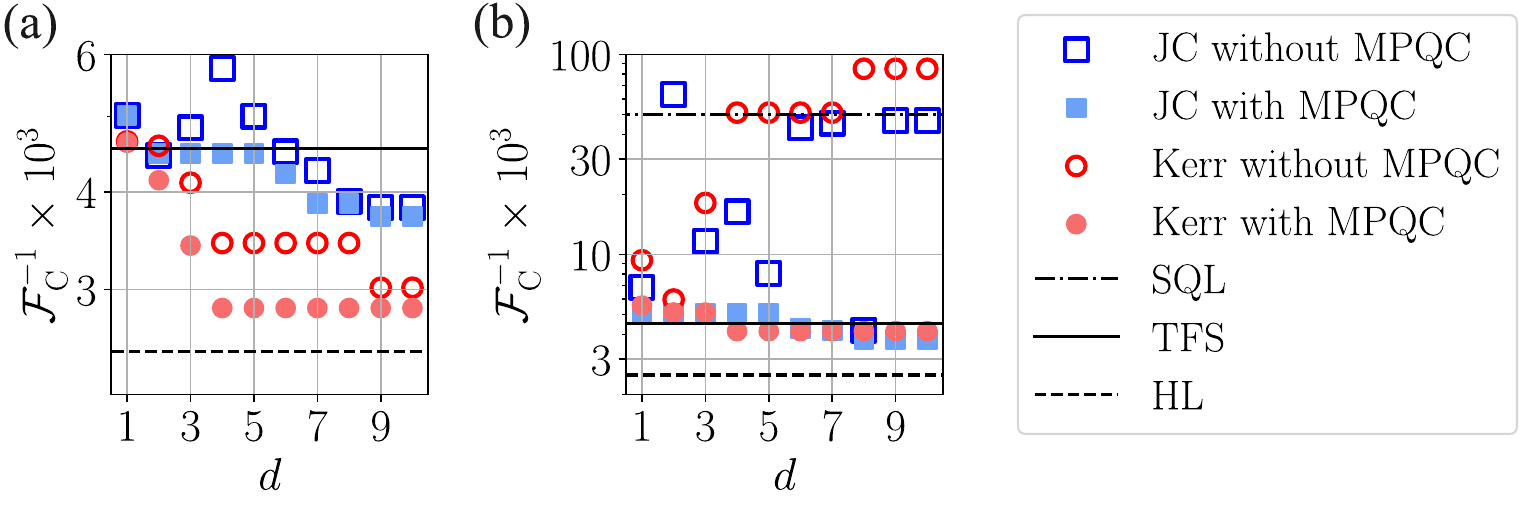}
    \caption{
    Inverse CFI $\mathcal{F}^{-1}_{\rm C}$ as a function of the number of layers $d$ obtained with the programmable approach using photon counting [panel (a)] and homodyne detection with a quadrature angle $\theta=0$ [panel (b)]. The mean-photon number is fixed at $N=20$. Solid markers represent the results obtained by employing the pre-measurement PQC  (labelled MPQC), while void markers are computed without the pre-measurement PQC. Blue squares (red circles) correspond to the JC (Kerr) ansatz. Dashed-dotted, solid, and dashed lines represent the standard quantum limit (SQL), the twin-Fock states (TFS) results, and the Heisenberg limit (HL), respectively.}
    \label{fig SM: measurement layers}
\end{figure*}

As we explain in Sec.~II of the Main Text, in the programmable approach we use of two different optimization circuits: The first one is aimed at preparing probe states featuring the largest possible QFI. The second one takes place immediately before the measurement and its objective is to maximize the CFI. Each optimization loop makes use of a different PQC, which we label the \textit{preparation} and the \textit{pre-measurement} PQCs, respectively. However, one may wonder whether it is necessary to perform such a second optimization step, since the probe state generated in the first optimization can already provide a large value of the CFI.

To assess this question, we examine the effect of the second optimization loop. In Fig.~\ref{fig SM: measurement layers} we plot the inverse of the CFI $\mathcal{F}^{-1}_{\rm C}$ obtained with and without the pre-measurement PQC. In the first case, the state after the Mach-Zehnder interferometer (MZI, see Sec.~II of the Main Text for more details) undergoes the second optimization loop using the pre-measurement PQC. This is the method used to compute the results shown in Sec.~V of the Main Text. However, in the second case, the pre-measurement PQC is absent and the CFI is directly computed using the output state of the MZI. We plot $\mathcal{F}^{-1}_{\rm C}$ as a function of the number of layers $d$ employed. When the pre-measurement PQC is present, both the preparation and the pre-measurement PQCs have $d$ layers. However, if the pre-measurement PQC is absent, $d$ is the number of layers of the preparation PQC.

Fig.~\ref{fig SM: measurement layers}(a) shows the results for photon-counting measurements. For both the JC and the Kerr ansätze, the values of $\mathcal{F}^{-1}_{\rm C}$ are always larger when the second optimization loop is not performed. This implies a worse metrological performance when the pre-measurement PQC is absent. However, as $d$ increases the two ansätze provide a larger metrological advantage than twin-Fock states (TFS) even in the absence of the second optimization loop. The necessity of the pre-measurement PQC is more evident for homodyne detection, as it is shown in Fig.~\ref{fig SM: measurement layers}(b). Here, the quadrature angle is fixed at $\theta=0$ (see Sec.~II of the Main Text). When one employs the two optimization loops, both the JC and the Kerr PQCs provide smaller values of $\mathcal{F}^{-1}_{\rm C}$ than TFS as $d$ increases. However, when the pre-measurement PQC is absent, for both ansätze $\mathcal{F}^{-1}_{\rm C}$ increases with $d$, even reaching the standard quantum limit (SQL). This implies that homodyne detection cannot provide any metrological advantage without the second optimization loop, even when one employs the probe states generated with the preparation PQC.

\begin{figure*}[tb]
    \centering
    \includegraphics[width=0.7\linewidth]{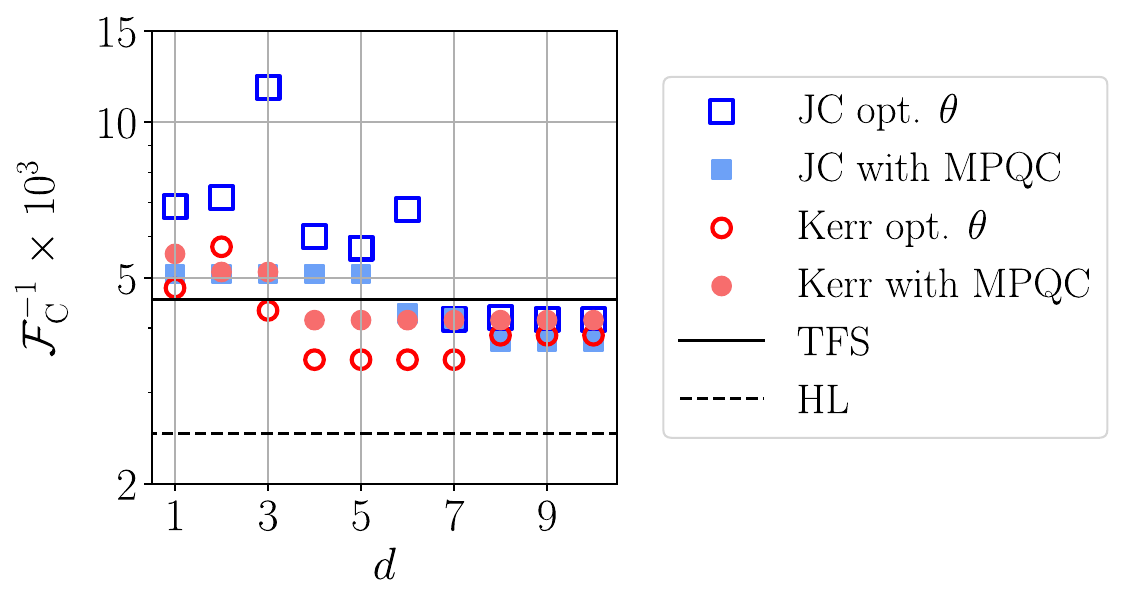}
    \caption{
    Inverse CFI $\mathcal{F}^{-1}_{\rm C}$ as a function of the number of layers $d$ obtained with the programmable approach using homodyne detection. The mean-photon number is fixed at $N=20$. Solid markers represent the results obtained by employing the pre-measurement PQC (labelled MPQC)  and a fixed value of the quadrature angle $\theta=0$. Void markers are computed by optimizing $\theta$ without the pre-measurement PQC. Blue squares (red circles) correspond to the JC (Kerr) ansatz. Solid and dashed lines represent the TFS results and the HL, respectively.}
    \label{fig SM: measurement layers opt theta}
\end{figure*}

However, it is possible to improve the results of homodyne detection without the pre-measurement PQC by optimizing the generalized quadrature angle $\theta$. In Fig.~\ref{fig SM: measurement layers opt theta} we compare the inverse CFI $\mathcal{F}^{-1}_{\rm C}$ obtained with the pre-measurement PQC and a fixed $\theta=0$ with the one attained by optimizing the value of $\theta$ to maximize the CFI without the pre-measurement PQC. We plot the resulting $\mathcal{F}^{-1}_{\rm C}$ as a function of the number of layers $d$ of the preparation PQC, which coincides with the number of layers of the pre-measurement PQC when this is included. For the JC ansatz, the values of $\mathcal{F}^{-1}_{\rm C}$ obtained by performing the optimization loop of the pre-measurement PQC with $\theta=0$ are still smaller than those obtained by optimizing $\theta$ without the pre-measurement PQC. However, especially for $d>6$, the difference between the values of $\mathcal{F}^{-1}_{\rm C}$ obtained with the two strategies is much smaller than when the quadrature angle is fixed to $\theta=0$ and the pre-measurement PQC is not employed (see Fig.~\ref{fig SM: measurement layers}). On the other hand, for the Kerr ansatz, the values of $\mathcal{F}^{-1}_{\rm C}$ obtained by optimizing $\theta$ without the pre-measurement PQC are smaller than those obtained by optimizing the pre-measurement PQC with $\theta=0$. Although the difference between the results of the two strategies is small, optimizing $\theta$ without including the pre-measurement PQC can be more efficient in terms of resources since optimizing a single parameter is faster than optimizing $3d$ or $2d$ of them, as required respectively by the JC and Kerr ansätze.

\section{Wigner distribution of the probe states generated with the programmable approach}
\label{sec SM: probe states programmable}

In this Section we analyze the probe states generated with the JC and Kerr ansätze in the programmable approach by studying their Wigner quasiprobability distribution. These are displayed in Fig.~\ref{fig SM: probe states programmable} for a fixed value of the mean-photon number $N=20$. For reference, in panel (a) [(d)] we plot the inverse of the QFI $\mathcal{F}^{-1}_{\rm Q}$ obtained using the JC (Kerr) ansatz as a function of the number of layers $d$ of the PQC. 

It is interesting to compare the Wigner quasiprobability distribution in phase space of the generated probe states for $d=1$ and for values of $d$ for which convergence is achieved ($d=8$ and $d=4$ for the JC and Kerr ansätze, respectively). As expected, for $d=1$ the states prepared by each ansatz [see panels (b) and (e)] belong to the same class as the states generated with the continuous approach at values of the interaction strength corresponding to the first minimum of $\mathcal{F}^{-1}_{\rm Q}$. In particular, for the JC ansatz a displaced cat state is produced, while the Kerr ansatz results in a $6$-component cat state, both featuring QFI values similar to those of twin-Fock states $\ket{N/2}\otimes\ket{N/2}$.

However, when the number of layers is increased, the output states of the two ansätze, featuring values of $\mathcal{F}^{-1}_{\rm Q}$ close to the Heisenberg limit (HL), become quite different from those produced with $d=1$. In panels (c) and (f), we plot the Wigner distribution of the states generated by the JC (Kerr) ansatz with $d=8$ ($d=4$), when the QFI has converged to its maximal value in each case. Interestingly, these states cannot be associated to any known class of states, which demonstrates the capability of the optimized PQC to generate non-trivial states.

\begin{figure*}[tb]
    \centering
    \includegraphics[width=\linewidth]{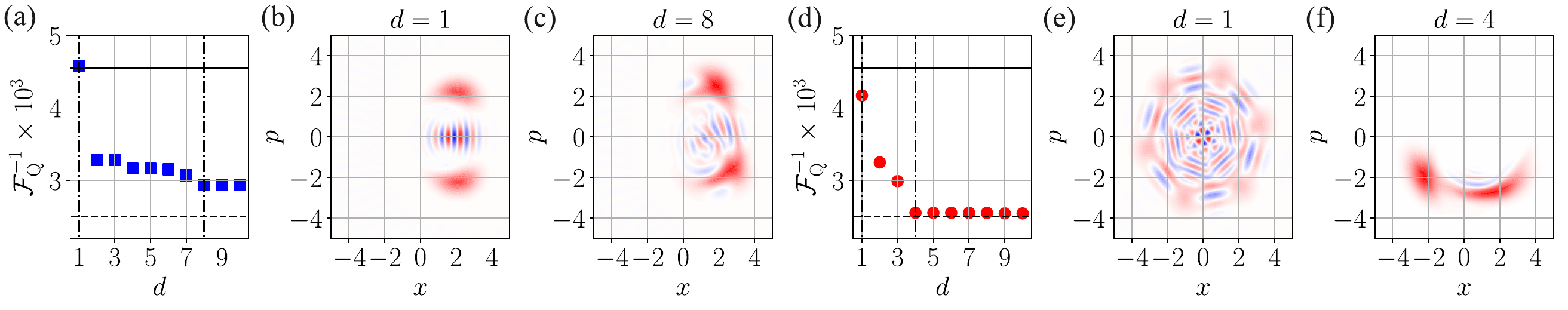}
    \caption{Generation of probe states in the programmable approach with a fixed mean-photon number $N=20$.
    (a) [(d)] Inverse QFI $\mathcal{F}^{-1}_{\rm Q}$ as a function of the number of layers $d$. Blue squares (red circles) are the results of the Jaynes-Cummings (Kerr) ansatz. Solid (dashed) lines correspond to TFS (the HL). Dashed-dotted lines signal the values of $d$ for which we plot the Wigner distribution of the output probe states of the parametrized quantum circuit. (b,c) [(e,f)] Wigner distribution in phase space of the states generated by the JC (Kerr) ansatz for $d=1$ and $d=8$ ($d=1$ and $d=4$).}
    \label{fig SM: probe states programmable}
\end{figure*}

\end{widetext}

\end{document}